# Machine Learning in Near-Field Communication for 6G: A Survey

Amjad Iqbal, Ala'a Al-Habashna, *Senior Member, IEEE*, Gabriel Wainer, *Senior Member, IEEE*, and Gary Boudreau, *Senior Member, IEEE*

*Abstract*— 6G wireless communication networks are expected to use extremely large-scale antenna arrays (ELAAs) to support higher throughput, massive connectivity, and improved system performance. ELAAs would fundamentally alter wave characteristics, transforming them from plane waves into spherical waves, thereby operating in the near field. Near-field communications (NFC) offer unique advantages to enhance system performance, but also present significant challenges in channel modeling, computational complexity, and beamforming design. The use of machine learning (ML) is emerging as a powerful approach to tackle such challenges and has the capabilities to enable intelligent, secure, and efficient 6G wireless communications. In this survey, we discuss ML-driven approaches for NFC. We first outline the fundamental concepts of NFC and ML. We then discuss ML applications in channel estimation, beamforming design, and security enhancement. We also highlight key challenges (e.g., data privacy and computational overhead). Finally, we discuss open issues and future directions to emphasize the role of advanced ML techniques in near-field system design.

*Index Terms*— 6G, extremely large-scale antenna arrays, machine learning, near-field communications

## I. INTRODUCTION

ACADEMIA and industry have defined a roadmap for the sixth-generation (6G) mobile communications, supporting emerging applications and services like immersive reality and the metaverse [1], [2], [3]. In June 2023, the International Telecommunication Union (ITU) released the recommended frameworks, timeline, and future technologies for 6G standardization [4], including three advancements: integrated artificial intelligence (AI) and communication, integrated sensing and communication (ISAC), and ubiquitous connectivity. They extended three use cased defined for 5G networks: immersive enhanced mobile broadband (eMBB), ultra-reliable low latency communication (URLLC), and massive machine-type communication (mMTC). Moreover, IMT-2030 presented customized key performance indicators for 6G, including nine enhanced and six new capabilities. 6G technology is expected to surpass 5G technology by a 100-fold increase in peak data rate (Tbps), 10-fold reduced latency (0.1ms) under a stringent reliability standard of 99.99999%, and 10-fold higher connection density [3], [5]. As 5G falls short of meeting these rigorous demands, there is a need to develop new technologies to address these challenges.

Some of the new technologies proposed to meet these stringent requirements include extremely large-scale (XL) antenna arrays (ELAAs), such as extremely large-scale multiple input multiple output (XL-MIMO), deployed at the base station (BS) to enhance capacity and coverage [6], [7], [8]. High frequencies (i.e., Terahertz (THz)) have the potential to provide abundant bandwidth resources [9], [10]. Additionally, metamaterials-based antennas exploit artificial materials to exhibit various desired electromagnetic (EM) characteristics for optimized wireless communications [11], [12]. As the dimensions and frequencies of the antenna increase smoothly, the effects of the near field become prominent, especially in XL-MIMO and THz communication networks. Therefore, understanding the role of near-field communications (NFC) in 6G networks is essential.

### A. The Role of NFC in 6G

The use of NFC in 6G networks can lead to a significant shift in EM wave characteristics. Typically, the EM field antenna can be divided into two distinct regions: the far-field and near-field regions. The EM wavefront can be propagated in the far-field region as a uniform planar wave. In the near-field region, the EM wavefront can be approximated as a non-uniform spherical wave due to the large array apertures and high operating frequency. The far-field and near-field boundaries can be determined using the Rayleigh distance $R_y = 2D^2/\lambda$ [13], where $D$ and $\lambda$ denotes the antenna apertures and transmission wavelength, respectively. In other words, if the distance between the BS and the user is greater than Rayleigh's distance, then the far-field propogation dominates. For near-field, the distance between BS and the user is less than the Rayleigh distance. As larger antenna arrays and higher frequency bands are deployed, the Rayleigh distance increases, expanding the

Amjad Iqbal, Ala'a Al-Habashna, and Gabriel Wainer are with the Department of Systems and Computer Engineering at Carleton University, Ottawa, Canada. Email: amjad.iqbal68a@gmail.com, alaaalhabashna@cunet.carleton.ca, gwainer@sce.carleton.ca
Ala'a Al-Habashna is also with the School of Computing and Informatics, Al Hussein Technical University, Amman, Jordan. Email: alaa.alhabashna@htu.edu.jo
Gary Boudreau is with Ericsson Canada, Kanata, Canada. Email: gary.boudreau@ericsson.com
This work is funded by Ericsson Canada and the Natural Sciences and Engineering Research Council of Canada (NSERC).

near-field region. For instance, [14] assumed a frequency of 2.4 GHz and developed a 3200-element antenna with an aperture size of $2m \times 3m$, resulting in a Rayleigh distance of $144m$, which is larger than the typical 5G cell radius. Consequently, a user within this distance experiences near-field propagation, making NFC an essential component of 6G networks. This needs a departure from conventional far-field models and the adoption of new techniques to effectively manage spherical wave propagation.

### B. The Need for Machine Learning in NFC

The transition from far-field to near-field presents several challenges, including designing efficient beamforming strategies to focus energy or power toward individual users and enabling high-precision user localization due to the high-dimensional data. Moreover, the presence of spherical wavefronts poses significant challenges in estimating accurate channel models in a highly complex environment. Using traditional approaches (i.e., alternating optimization (AO), heuristics, and dynamic programming) requires enormous computational capabilities due to the large number of antennas and intricate user interaction in near-field scenarios. To address these challenges, machine learning (ML) has emerged as a powerful tool to leverage data-driven models and optimize beamforming strategies, learn complex channel characteristics, and enhance the reliability and efficiency of wireless communication. Advanced ML techniques have significantly reduced the computational overhead and improved the efficiency of near-field beam training, making real-time adaptation feasible. Furthermore, ML-based localization algorithms can exploit near-field propagation effects to achieve sub-centimeter accuracy, a critical application requirement for autonomous systems and the Internet of Everything (IoE). In the following sections, we will examine the role of ML in near-field beamforming, localization, and computational optimization.

### C. Motivations, Contributions, and Structure

The early generations of wireless communication (i.e., 1G to 5G) were typically based on low operating frequencies and small-scale antenna arrays, making the near-field effect negligible. Thus, the related literature work on near-field was very sparse. However, with the emergence of 6G and the increasing deployment of ELAA's and THz, the near-field effect has become significantly more prominent, providing opportunities for in-depth exploration of the near-field. Although some existing literature (e.g., [13], [15], [16], [17]) provides a high-level introduction to NFC, it lacks a fundamental understanding of its applications. Moreover, some of the related work (e.g., [18], [19], [20], [21]) do not consider the role of ML in the near-field, while others (e.g., [22], [23], [24]) have considered the ML but not explore its application comprehensively. Therefore, in this survey, we address these gaps by focusing on integrating the role of ML in NFC for 6G wireless networks, emphasizing its applications, challenges, and future directions. With the advent of ELAA, the nature of the wireless communication wavefront has shifted from far-field to near-field regimes, offering new challenges and opportunities for efficient communication. Using traditional communication models is insufficient for handling the complex wave behaviors introduced by ELAA, requiring new approaches for efficient system design. ML provides a promising solution by leveraging data-driven techniques to enhance system performance. Nonetheless, ML introduces its own challenges, such as computational complexity, scalability, and privacy concerns. Thus, in this paper, we aim to outline the potential role of ML in NFC, outline its limitations, and propose future research directions to build intelligent and efficient 6G communication systems. Therefore, in this paper, we:

● Introduce a comprehensive analysis of the fundamental differences between far-field and NFC, emphasizing their distinct wave characteristics. We explain the boundary separating these two regimes and extend its classical derivation to include reconfigurable intelligent surface (RIS-aided MIMO) networks. This provides new insights into the transition from traditional wireless communication models to the emerging near-field paradigm, laying the foundation for further exploration of advanced techniques in this area.

● Systematically review the role of the ML-based approach in NFC and categorize it into two groups based on their objectives. The challenges associated with the near-field (channel modeling, channel estimation, beamforming), where conventional techniques often struggle due to the wavefront curvature and increased spatial complexity, are addressed in the first category. The ML techniques that aim to exploit the advantages of NFC (e.g., improved spatial multiplexing and channel capacity enhancement) are highlighted in the second category. In this context, ML techniques are used to optimize system performance through intelligent user scheduling, resource allocation, and transmission strategy design. We present a clear framework for how ML can effectively enhance key aspects of near-field wireless communication.

● Showcase the results of several case studies demonstrating the effectiveness of ML-driven approaches in NFC. These case studies demonstrate how ML techniques can significantly enhance channel estimation, beamforming design, and security in practical scenarios. Finally, we identify open challenges and future research directions, including optimizing the Rayleigh distance with respect to different communication metrics, developing computationally efficient ML models for real-time applications, and exploring hybrid near-field and far-field communications (FFC) strategies. Additionally, we discuss the potential of advanced ML techniques in shaping the future of NFC systems. This survey aims to serve as a foundation for future research in this field, highlighting both the opportunities and challenges associated with integrating ML with NFC.

The structure of the paper is designed to provide a logical and comprehensive flow of information, beginning with foundational concepts and gradually exploring advanced ML applications, challenges, and future research directions in NFC for 6G networks.

Section II presents the theoretical background of NFC, propagation characteristics, and the key differences from FFC.

Table I. List of Acronyms

| Acronyms | Definitions |
| --- | --- |
| 5G | Fifth-generation |
| 6G | Sixth-generation |
| AD | Anomaly detection |
| AE | Autoencoders |
| AI | Artificial intelligence |
| AL | Adversarial learning |
| AO | Alternating optimization |
| AR | augmented reality |
| BER | Bit error rate |
| BS | Base station |
| CNNs | Convolutional neural networks |
| CR | Cognitive radio |
| DSA | Dynamic spectrum allocation |
| CSI | Channel state information |
| DL | Deep learning |
| DNN | Deep neural network |
| DP | Dynamic programming |
| DRL | Deep reinforcement learning |
| ELAAs | Extremely large-scale antenna arrays |
| EM | Electromagnetic |
| eMBB | Enhanced mobile communication |
| FFC | Far-field communication |
| FL | Federated learning |
| GA | Genetic algorithm |
| GANs | Generative adversarial networks |
| GPUs | Graphic processing units |
| IoE | Internet of Everything |
| IS | Interference suppression |
| ISAC | Integrated sensing and communication |
| ITU | International Telecommunication Union |
| KPI | Key performance indicators |
| LSTM | Long short-term memory |
| MARL | Multi-agent reinforcement learning |
| MIMO | Multiple-input multiple-output |
| ML | Machine learning |
| MMSE | Minimum mean square error |
| mMTC | Massive machine-type communication |
| NFC | Near-field communication |
| RA | Resource allocation |
| RIS | Reconfigurable intelligent surface |
| RL | Reinforcement learning |
| RNNs | Recurrent neural networks |
| SA | Search algorithm |
| SL | Supervised learning |
| SNR | Signal-to-noise ratio |
| SVM | Support vector machine |
| THz | Terahertz |
| TL | Transfer learning |
| TPUs | Tensor processing units |
| UE | User Equipment |
| UL | Unsupervised learning |
| UPW | uniform planar wave |
| URLLC | Ultra-reliable low-latency communication |
| VLC | Visible light communication |
| WPS | Wireless power transfer |
| XL | Extremely large-scale |
| ZF | Zero forcing |

Section III introduces ML and relevant uses of ML for NFC. Section IV shows the application of ML in NFC within areas such as security and privacy enhancement, energy-efficient communication, interference management, and spectrum optimization. Section V discusses several case studies to showcase the advances of ML in NFC. The challenges, limitations, and future research directions are highlighted in

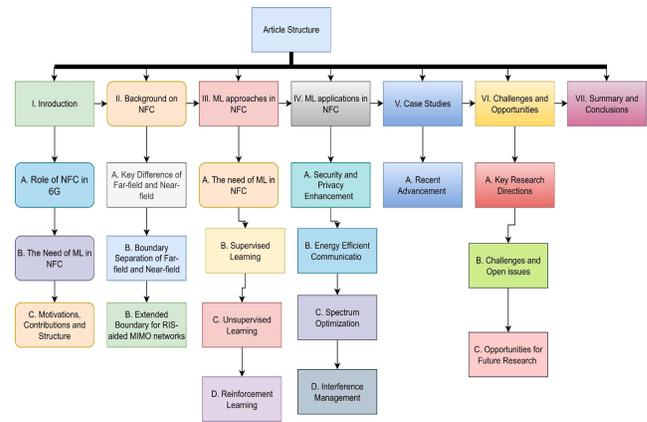

Fig. 1. Article Structure

Section VI. A summary of the paper is presented in Section VII. The definitions of the main acronyms used in this paper are summarized in Table I, for ease of reference, whereas the structure of the paper is illustrated in Fig. 1.

## II. BACKGROUND ON NEAR-FIELD COMMUNICATION

The past three decades have witnessed a transformative shift in the telecommunications sector, driven by the adoption of extremely high frequencies, the development of massive antenna arrays, and the emergence of new metamaterial-based antennas. This shift not only enables the network capacity and efficiency of wireless communication but also reshapes the nature of EM radiating characteristics.

The EM radiation waves from an antenna can be categorized into two regions: the far-field and near-field regions [25], [26]. Each region exhibits a distinct propagation behavior, which influences the beamforming strategies, signal transmission, and overall communication performance. The EM wavefront is approximately planar in the far field, meaning the waves travel in parallel, making traditional beamforming techniques (e.g., a phased array) more effective. However, the wavefront has a complex shape in the near-field due to the reduced distance between the BS and the user. The difference between far-field and near-field is visualized in Fig. 2 [27]. The Rayleigh distance can be used to distinguish these two regions [28], [29].

Near-field was limited to a small spatial range in earlier generations, within a few centimeters to a few meters, due to low operating frequencies and low-dimensional antennas. Therefore, the importance of the near-field was negligible in such a scenario. Thus, the far-field approximation is designed to simplify wave propagation for these technologies efficiently.

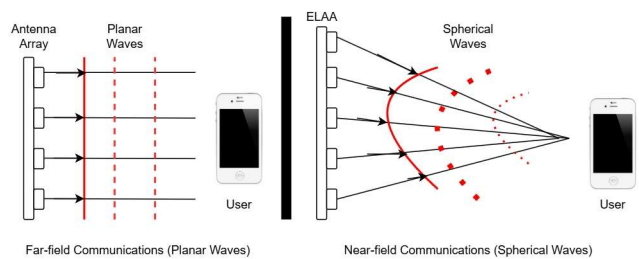

Fig. 2. FFC vs NFC

The advent of ELAA and the use of extremely high frequencies (i.e., THz) in future wireless communication networks (i.e., 6G) have changed the nature of the problem. Extremely high frequencies have much shorter wavelengths, which pushes the boundaries of near-field regions from a few meters to potentially hundreds of meters. Moreover, using metamaterial-based antennas, which enable dynamic wavefront shaping and ultra-high spatial resolution, further complicates wave propagation, making far-field approximations inadequate for accurate system modeling and performance prediction [30], [31]. Such a drastic expansion of the near-field region challenges traditional far-field-based communication models, necessitating a shift toward a new NFC paradigm.

The transformation underscores the significant importance of near-field effects in 6G networks. This enables the precise manipulation and control of EM waves in an extended near-field region, unlocking new opportunities for novel applications. Moreover, NFC enables spatially selective energy-focusing, enhancing spectral efficiency and reducing interference. Furthermore, NFC is pivotal for near-field wave propagation in emerging technologies, such as RIS, wireless power transfer (WPS), and sub-wavelength localization.

Given these advancements, the importance of near-field is not negligible in 6G networks, which motivates the investigation of the new NFC paradigm. To shed light on the benefits of NFC, we first distinguish between near-field and far-field in terms of their EM-radiation properties.

### A. Key Difference between Near-field and Far-field

According to antenna theory, the EM field surrounding the transmitter (BS) and receiver (user) can be divided into far-field and near-field regions [32]. The near-field region can be further separated into two subregions (i.e., reactive near-field and radiating near-field). The boundaries for these regions can be approximated by the operating wavelength and physical antenna dimension (apertures). Therefore, understanding these regions is essential for modern wireless communication. We explain these three regions in the following:

- The *reactive near-field* is limited to the closest region between the transmitter and receiver and can be extended up to approximately $0.62\sqrt{D^3/\lambda}$. The EM fields within this region are predominantly reactive, meaning the antenna can release and store energy rather than propagating outward as radiating waves. This energy release results in strong evanescent waves, which decay exponentially with distance and do not contribute significantly to long-range wireless communication. The fields in this region do not effectively radiate and are generally impractical for traditional wireless communication over long distances.

- The *radiating near-field* can be located for a distance from transmitter to receiver that is greater than a few wavelengths and can be approximated up to $2\,D^2/\lambda$. The wavefronts exhibit a noticeable spherical curvature in this region, and the phase variations across the antenna aperture are significant. In this region, the radiation pattern follows the angular field distribution and is dependent on both distance and angle. This region is vital in modern wireless technologies, where beamforming and focusing are needed for efficient communication. Advanced signal processing techniques are often required in the radiating near-field region to manage beam steering and wavefront shaping effectively.

- The *far-field* can be located from transmitter to receiver at distances greater than $2\,D^2/\lambda$. This region is also known as the Fraunhofer region. In the far-field, the wavefronts become nearly planar, meaning the relationship of the phase and amplitude of the waves remains stable over increasing distances. The angular field distribution of the far field is independent of distance, making it suitable for traditional antenna analysis and radiation pattern measurements.

To summarize these three regions, the reactive near-field is typically small, where the evanescent waves exponentially decay with distance. The radiating near-field generally occurs from a few meters to hundreds of meters, where the spherical wavefront dominates this region. The far-field typically occurs for a greater distance, where the nature of EM waves becomes planar. The propagation nature of these three regions is depicted in Fig. 3. For the remainder of the paper, we will discuss the radiating near-field (for simplicity, referred to as the near-field) and far-field regions.

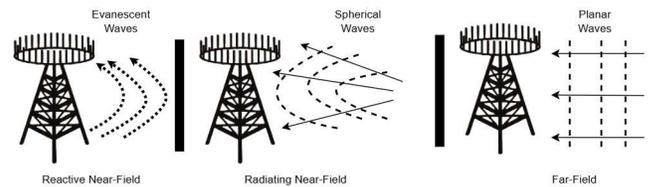

Fig. 3. Propagation nature of EM waves

### B. Boundary Separation of Far-Field and Near-Field

There is no strict boundary separation between the near-field and far-field areas, as the transition between these regions occurs gradually. Therefore, different studies introduce distinct metrics to identify the field boundary areas. The channel gain error perspective and phase error perspective are the two widely used metrics to define these measurement criteria.

- *Channel gain error perspective*: In this case, a more accurate description of the channel gain becomes possible for the off-axis region of the field boundary. Specifically, based on the Friis formula, the channel gain is inversely proportional to the square of the distance [33]. This approximation is not valid in the near-field region due to the presence of complex spherical waves. The far-field region can be established through the Friis formula approximation of the actual channel gain with a tolerable error margin. The field boundary depends on aperture size and wavelength and requires consideration of departure angle, arrival angle, and transmit antenna aperture shape, according to this perspective.

- *Phase error* perspective: In this case, several rules of thumb are commonly used, such as the Fraunhofer condition [34], the Rayleigh distance [35], and the extended Rayleigh distance for MIMO transceivers and RISs [36] to define the field boundary. These distances primarily apply to the field boundary near the main axis of the antenna aperture, with less applicability to off-axis regions.

### C. Extending Boundary Separation for RIS-aided Network

In classical wireless communication, the Fraunhofer and Rayleigh distances traditionally define the transition between far-field and near-field regions. The far-field region follows the Friis free-space transmission $P_r = P_t G_t G_r \lambda^2/(4\pi d)^2$, where wavefronts can be approximated as planar, and holds for distances beyond the Fraunhofer distance $d_F = 2D^2/\lambda$. However, the classical Friis model does not hold in the near-field region due to significant phase variations where wavefronts become complex and spherical. Rather than using conventional far-field assumptions, the Rayleigh distance $d_R = 2D^2/\lambda$ serves as a transition boundary for the near-field. This transition becomes particularly relevant in emerging technologies, such as RIS-aided MIMO systems, where near-field effects play a crucial role in performance improvement. Given that an RIS implementation consists of a large number of passive reflecting elements, its use enables dynamic wavefront manipulation, effectively altering the EM propagation environment. Unlike conventional relay systems, where signals undergo amplification and retransmission, RIS passively reflects incoming signals with controlled phase shifts, introducing new wave transformation effects that extend the near-field region. The presence of an RIS alters the adequate aperture size and redefines the boundaries for near-field and far-field regions. In particular, when an RIS is introduced between the transmitter and receiver, the effective Rayleigh distance must account for the RIS aperture size $d_{RIS}$, leading to a modified near-field boundary expressed as $d_{RIS} = 2(D_T + D_{RIS} + D_R)^2/\lambda$, where $D_T$ and $D_R$ shows the effective transmitter and receiver apertures, respectively. This extension suggests that the near-field region extends significantly beyond classical Rayleigh predictions for large RIS deployments, enabling new spatial focusing capabilities. In contrast to the conventional far-field, where beamforming techniques rely on angular steering, RIS-assisted near-field MIMO exploits spatial focusing, where signals are concentrated on specific locations rather than just specific angles, fundamentally changing how signal transmission is optimized. Research has recently been performed to support the theoretical and practical concept of the near-field effects in RIS-aided MIMO networks. For instance, [37] proposed a simultaneously transmitting and receiving RIS-aided MIMO framework, focusing on controlling wavefront curvature to enhance the weighted rate precisely. In [38] and [39], piece-wise channel models were proposed to characterize the near-field propagation in RIS-aided MIMO networks. The power scaling is analyzed in RIS-aided systems [40]. Moreover, [41] proposes a wideband beamforming scheme for RIS-aided near-field MIMO systems to mitigate the low beamforming performance caused by the near-field double beam-split effect. These works [37]-[41] emphasize the importance of RIS in enhancing NFC by reshaping the traditional electromagnetic boundary conditions. Fig. 4 illustrates the extended conventional Rayleigh distance derived in an RIS-aided network. The cascaded channel (BS-RIS-UE) links are considered, where the distance of BS-RIS ($r_1$) and RIS-UE ($r_2$) are added to calculate the phase discrepancy. Moreover, the harmonic mean is used to determine the range for near-field in an RIS-aided network (i.e., $r_1 r_2/r_1 + r_2 < 2 D_{RIS}^2/\lambda$) while keeping the most considerable phase discrepancy of $\pi/8$. As shown in Fig. 4, when the harmonic distance is shorter than the Rayleigh distance, the near-field dominates the RIS-aided network, reinforcing the importance of near-field effects in future wireless systems.

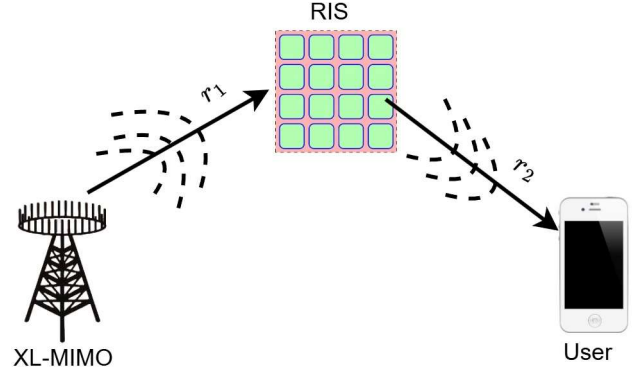

Fig. 4. RIS-aided near-field MIMO network.

### III. MACHINE LEARNING APPROACHES IN NFC

NFC plays a key role in future wireless communication systems, particularly in 6G networks, where ELAA and high-frequency bands are utilized. NFC operates over shorter distances between transmitter and receiver, where the behavior of the EM wavefront is spherical rather than planar. This unique characteristic presents significant challenges, including precise beamforming, complex channel modeling, and accurate signal detection, all of which are critical for achieving high data rates, low latency, and reliable connectivity. Conventional approaches, such as AO [42], zero-forcing (ZF) [43], minimum mean square error (MMSE) [44], genetic algorithm (GA) [45], and search algorithm (SA) [46], [47], are well-established approaches to tackle these challenges effectively; however, these approaches face significant challenges in modern NFC systems, especially in dynamic and complex environments.

The primary issue with these approaches stems from the inherent complexity of near-field channels. Moreover, these approaches use analytical models based on EM theory to characterize these channels, such as Green's function [23], [48], and ray tracing [49]. These approaches are well-suited for simple and static environmental scenarios; however, they often fail to capture the intricate spatial and temporal variations in real-world near-field environment problems. For instance, the spherical wavefront in NFC leads to rapid changes in channel characteristics over short distances, making it challenging to model accurately using conventional methods [50]. Furthermore, these traditional approaches require iterative optimization, which is computationally intensive and lacks the adaptability needed for real-time applications [45]. Similarly, classical signal detection algorithms, like maximum likelihood detection, struggle to perform well in noisy and interference-prone near-field environments [51].

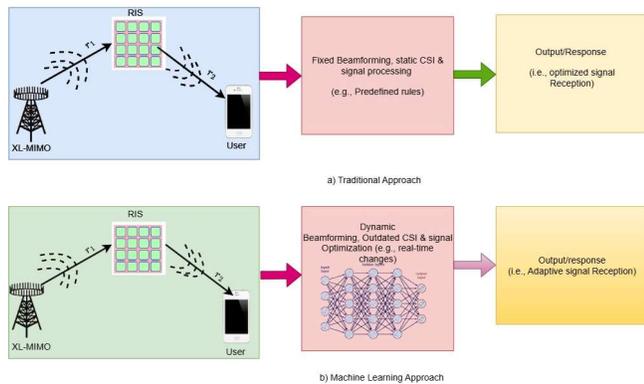

Fig. 5. Traditional approaches vs Machine Learning

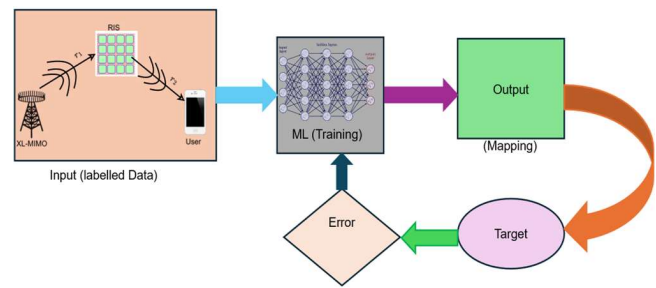

Fig. 6. Supervised Learning Approach

ML is a more powerful and effective solution to address these limitations for NFC [52]. ML techniques directly learn complex patterns and relationships from data without explicitly relying on mathematical modeling. This data-driven approach enables ML to handle the intricacies of near-field channels more effectively. For example, ML models can learn the mapping between pilot signals and channel states, achieving higher accuracy in channel estimation than traditional approaches [53]. Similarly, ML-based beamforming algorithms can optimize beamforming vectors in real time, adapting to changes in the environment without the need for computationally expensive iterations [54]. Moreover, ML techniques, such as deep learning (DL), excel in tasks like signal detection by learning complex decision boundaries in noisy environments, outperforming classical algorithms in terms of accuracy and robustness [55]. Fig. 5 explicitly illustrates the difference between traditional approaches and ML for RIS-aided MIMO networks.

Several factors drive the transition of ML to NFC. Firstly, the availability of large datasets generated through simulations or real-world measurements provides the necessary training data for ML models. Secondly, advancements in computational hardware, such as graphic processing units (GPUs) and tensor processing units (TPUs), enable the efficient training and deployment of complex ML models. Thirdly, the need for scalable and adaptive solutions in next-generation communication systems, such as 6G, has further accelerated the adoption of ML in NFC [56]. Leveraging ML in NFC networks not only boosts the performance of future wireless communication networks but also achieves higher performance, greater adaptability, and improved scalability, paving the way for innovative applications in areas like RIS, MIMO, dynamic RA, and precise localization.

ML offers diverse techniques to address the challenges in NFC, each tailored to specific tasks such as channel estimation, beamforming, signal detection, and user localization. One of the most widely used approaches is supervised learning (SL) [57], which excels in tasks where labeled data is available. For instance, in channel estimation, SL models such as K-nearest neighbor and support vector machines (SVMs) are trained on datasets containing input-output pairs, where the input is typically pilot signals and the output is the corresponding channel state. These models learn the mapping between the input and output, enabling accurate channel estimation even in near-field environments. Similarly, SL can be applied to beamforming optimization, where models are trained on predefined datasets containing optimal beamforming solutions. This approach reduces the computational overhead of traditional iterative optimization methods and achieves high accuracy for a large dataset environment scenario. However, SL is effective where pre-defined datasets are available (i.e., complete information about the characteristics of CSI and beamforming), which is difficult to obtain in real-time scenarios. The basic structure of SL is depicted in Fig. 6.

Another powerful ML approach is unsupervised learning (UL), which is particularly useful where the labeled data's priority is no longer needed [58]. For example, in user localization, UL techniques like K-means clustering can group users based on their spatial locations in near-field scenarios. This is especially valuable in multi-user environments, where precise localization is critical for efficient RA and interference management. Additionally, UL techniques such as principal component analysis (PCA) can be employed for dimensionality reduction, thereby simplifying the complexity of near-field channel data and enhancing the efficiency of subsequent processing tasks, including beamforming and signal detection. The UL approach provides a good solution for handling large datasets in NFC without relying on labeled data, offering advantages such as user clustering and dimensionality reduction. However, UL often fails to capture the complex, non-linear characteristics of near-field channels, making it less suitable for precise and adaptive network optimization. An illustrating diagram of UL is shown in Fig. 7.

Reinforcement learning (RL) is a widely used ML approach to optimize 5G and beyond wireless communication systems [59], [60], [61] and has gained a lot of traction in NFC due to its ability to handle dynamic optimization tasks [62]-[63]. In RL, the agent learns to find optimal policies through trial and error, adapting to changes in the environment in real-time. For instance, RL can be used for dynamic beamforming, where the algorithm continuously interacts with the environment to optimize beamforming strategies based on feedback in the form of rewards (e.g., signal-to-noise ratio (SNR)). Similarly, RL is well-suited for RA, where it can optimize power and spectrum allocation by learning policies that maximize system performance while minimizing resource consumption [64]. Although RL can handle large state and action space problems, it is typically well-suited to solve problems with discrete state and action spaces. One of the main disadvantages of RL is that state and action values are stored in a tabular form, which makes it challenging for RL to scale large and complex environmental problems [65]. Despite such challenges, RL has shown promising results in modern problems. However, in wireless communication, selecting RL is crucial to consider its applicability to the specific problem. RL can be a powerful tool in the context of the NFC optimization problem, but its effectiveness depends on the problem's complexity and nature. The basic structure of RL is shown in Fig. 8.

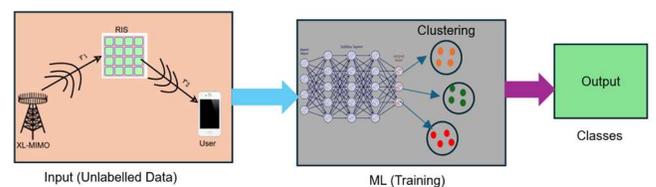

Fig. 7. Unsupervised ML approach

To address complexity and high dimensionality, a deep neural network (DNN) is integrated into RL, resulting in deep reinforcement learning (DRL). The DNN represents the policy and value function for DRL. This enables the DRL models to learn more complex and abstract representations of the state and action spaces, allowing the DRL model to handle complex environments effectively that can be too large or complicated for traditional RL techniques. By leveraging the power of DNNs, DRL can achieve better performance and more efficient exploration of the environment, making it particularly well-suited for challenging tasks in NFC.

In NFC, where distances between transmitters and receivers are significantly reduced, the EM wave propagation becomes highly sensitive to small changes in position, orientation, and surrounding objects. In such scenarios, DRL has the ability to learn and adapt to rapidly changing conditions. This is because DRL does not require prior information about the environment. For example, DRL can continuously optimize beamforming vectors in dynamic beamforming by learning from real-time feedback, such as SNR or CSI. This is particularly useful in NFC, where the near-field region requires precise control over beamforming to account for spherical wavefronts and rapid signal decay.

Moreover, DRL is effective in RA tasks, where it can dynamically adjust power and spectrum usage based on the current environmental conditions. In NFC, where interference and user demand changes frequently, DRL can learn policies that maximize system performance while minimizing resource consumption. For example, in a densely populated NFC environment, DRL can allocate resources to users to balance fairness and efficiency, adapting to changes in user mobility and channel conditions.

The ability of DRL to generalize from past experiences and explore high-dimensional spaces makes it a powerful tool for solving real-world NFC problems. However, DRL also introduces new challenges, such as increased computational complexity and the need for large amounts of training data. Techniques such as experience replay, target networks, and transfer learning (TL) are often employed to enhance the stability and efficiency of DRL training. Despite these challenges, DRL has demonstrated significant potential in NFC applications, offering a scalable and adaptive solution for optimizing wireless communication systems.

## IV. APPLICATIONS OF ML IN NFC

NFC is a transformative technology for future wireless communication networks, especially in scenarios where the distance between the transmitter and receiver is within the Rayleigh distance [66]. In NFC, the EM wavefront exhibits unique propagation characteristics that differ significantly from those of FFC (i.e., spherical instead of planar). These characteristics, including precise spatial correlation and phase variations, present new opportunities for optimizing NFC systems. However, they also open up complexities that traditional methods struggle to address. ML emerges as a powerful tool to harness these opportunities, offering intelligent and adaptive solutions for enhancing NFC performance. This section explores ML applications in NFC, focusing on four key areas: *security and privacy enhancement*, *spectrum optimization*, *interference management*, and *energy-efficient communication*. We highlight how ML leverages NFC in those areas to achieve superior performance, particularly for the BS-RIS-user link scenarios.

### A. Enhancing Security and Privacy

Security and privacy are always the prime objectives of any wireless communication network. The short-range communication inherent in NFC systems consistently poses considerable security risks, primarily due to high user density, increased interference, vulnerability to jamming attacks, and potential eavesdropping by nearby adversaries. Moreover, future wireless communication networks are based on ELAA, and threats are becoming more sophisticated, including the use of spoofing, relay attacks, and physical layer security breaches. ML has emerged as a more powerful tool to tackle these challenges by leveraging near-field channel characteristics for enhanced security issues. This subsection explores ML applications in NFC, focusing on anomaly detection, channel fingerprinting, adversarial learning, and physical layer key generation to address security and privacy issues.

### 1) Anomaly Detection in NFC

Anomaly detection (AD) enables the identification of unauthorized access and is a critical component of NFC security to detect attempts at eavesdropping or spoofing. The near-field channel exhibits unique signal characteristics, such as phase, amplitude, and spatial correlation, which can be analyzed to detect deviations from normal behavior. ML models, particularly SVMs and DNNs, have been widely adopted for detection purposes. DNNs are trained on large datasets of standard signal patterns and their corresponding CSI. During inference, the proposed model compares real-time signals with learned patterns to identify potential threat deviations. The proposed approach demonstrates an accuracy of over 99% in reporting detection rates in NFC [67].

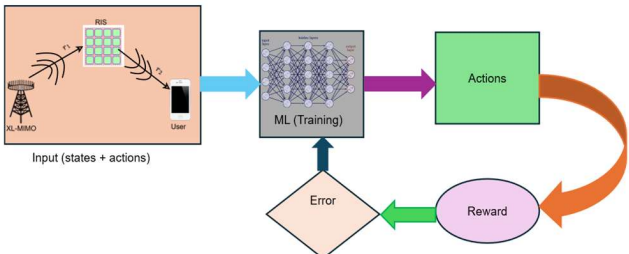

Fig. 8. Reinforcement Learning.

Besides SL, UL techniques such as autoencoders (AE) have shown promising solutions in detecting novel attack patterns. AE learns to reconstruct standard signal patterns; any significant reconstruction error indicates an anomaly. This approach is particularly useful in cases where dynamic environments lack labeled attack data. The AE approach has been demonstrated to achieve a false positive rate of less than 1.2% for detecting zero-day attacks [68]. These ML-based AD systems provide a robust defense against physical layer attacks, ensuring the integrity and confidentiality of NFC communications.

### 2) Channel Fingerprinting for Device Authentication

Channel fingerprinting can exploit the unique characteristics of near-field channels, such as spatial correlation and phase shifts, to create distinctive fingerprints for devices and legitimate users. These fingerprints can be used for real-time device authentication, preventing spoofing attacks and unauthorized access. ML models, especially convolutional neural networks (CNNs), have been widely employed to classify these fingerprints accurately. For example, a CNN model is trained to map the spatial correlation of the channel to a specific device or user, enabling robust authentication even in dynamic scenarios.

Several researchers have demonstrated the effectiveness of channel fingerprinting in recent years. For example, a CNN-based framework has been shown to achieve a false acceptance rate (FAR) of less than 0.01 [69]. Additionally, the use of RL enables the real-time adaptation of the fingerprinting model in wireless communication scenarios, where channel conditions can change constantly. This approach ensures continuous authentication, maintaining an accuracy of over 23.13% even under varying environmental conditions [70]. By combining the spatial sensitivity of near-field channels with the adaptive capabilities of ML, channel fingerprinting provides a powerful mechanism for enhancing NFC security.

### 3) Adversarial Learning for Threat Simulation

Recently, adversarial learning (AL), particularly generative adversarial networks (GANs) [71], has attracted a lot of attention and has emerged as a powerful tool in NFC systems for simulating and mitigating security threats. GAN comprises two neural networks (i.e., a generator and a discriminator). For threat simulation, the generator creates synthetic attack data (e.g., interfered signals), whereas the discriminator distinguishes between real and synthetic data. ML models can learn to recognize and respond to real-world threats, such as jamming or interfering with attacks, by training on synthetic attack data. For instance, a GAN is capable of simulating jamming attacks and developing countermeasures, such as signal masking or frequency hopping, for NFC systems [72]. This helps to offer an intelligent strategy for jamming and reduces the success rate of jamming attacks in communication systems. Moreover, AL can stabilize ML models against adversarial attacks, where an attacker attempts to manipulate the input data to deceive the model. In [73] adversarial training approach is employed to reduce the vulnerability of ML models. The work above highlights the potential advancements of adversarial learning to enhance the robustness of NFC security systems.

### 4) Physical Layer Key Generation

Physical layer key generation leverages the inherent randomness and unique characteristics of the wireless communication channel to generate a secret key for secure communication in near-field channels. This approach has been highly effective in rapidly spatially decorrelating the near-field channel, particularly for short distances, to generate high-entropy keys. ML approaches, such as RL and DL, have been widely employed to optimize the key generation process. For example, RL is proposed to minimize the bit error rate (BER) while maximizing the entropy of the generated keys [74]. This is achieved by dynamically adjusting the sampling rate or quantization levels based on the channel conditions. This RL-based framework is demonstrated to achieve a BER of less than $10^{-6}$ with a key generation rate of 10 bits per second (bps) [75]. Moreover, DL models have also been used to predict channel variations and generate keys in dynamic environments. For example, a DL model is proposed to create keys with an entropy of over 0.95, ensuring high security for NFC systems [76]. These ML-based key generation techniques provide a scalable and efficient solution for securing NFC communications in 6G networks.

### B. Spectrum Optimization

NFC introduces opportunities and challenges for spectrum optimization due to the dynamic characteristics of near-field EM propagation, including the variation in spatial correlation and phase. ML emerged as a powerful tool to tackle the spectrum utilization challenges, enabling adaptive and intelligent solutions for efficient spectrum utilization in NFC systems. This subsection explores three key ML-driven NFC spectrum optimization applications: dynamic spectrum allocation (DSA), spectrum sensing and classification, and cognitive radio (CR) networks.

### 1) Dynamic Spectrum Allocation

DSA refers to the demand for spectrum resources, which varies significantly due to the high density and short range of NFC, and plays a crucial role in NFC systems. RL emerged as a powerful ML algorithm that can be used for DSA to enable adaptive and real-time decision-making. RL algorithms (i.e., Q-learning and deep Q-networks) can find the optimal strategies for spectrum allocation by interacting with the NFC environment. For instance, an RL agent can allocate spectrum resources to NFC devices based on real-time demand and environmental conditions, maximizing throughput while minimizing interference and latency [77]. The unique spatial correlation of near-field EM waves allows RL models to optimize spectrum allocation for specific user locations and device configurations [78]. Moreover, to handle the multi-user NFC environmental problem, multi-agent reinforcement learning (MARL) can enable collaborative spectrum allocation among multiple devices. In such a scenario, each agent (e.g., BS, RIS, or a user device) learns to coordinate its spectrum usage with other agents, ensuring efficient resource sharing and minimizing obstacles. MARL-based approaches have achieved near-optimal spectrum utilization in dynamic and heterogeneous NFC environments [79].

### 2) Spectrum Sensing and Classification

Spectrum sensing is a technique used in wireless communication to detect unused portions of the radio frequency (RF) spectrum [80], [81], [82]. Spectrum sensing is used to identify and detect available frequency bands within the near-field region for NFC. The spherical wavefront and precise spatial correlation of EM waves in NFC create unique spectral signatures that can be leveraged for accurate spectrum sensing. DL models have proven to be an effective approach for spectrum sensing. For example, CNNs are well-suited for analyzing spectral data in NFC due to their ability to capture spatial and temporal patterns. The CNN models can process power spectral density measurements from near-field EM waves to detect unused or underutilized frequency bands. After training, these models have the capability to classify vacant and occupied bands with high accuracy, even in the presence of noisy channels [83]. This is particularly valuable for precisely localizing devices that emit near-field EM waves, enabling fine-grained spectrum sensing. This work is further extended by utilizing TL techniques to allow the fine-tuning of pre-trained CNN models for specific NFC environments. For this reason, a pre-trained CNN model based on the spectral data is adapted to near-field scenarios by retraining the final layers of the network. The results indicate that large labeled datasets are no longer needed and accelerate the deployment of NFC systems [84].

### 3) Cognitive Radio Networks

CR networks enable NFC devices to adapt and sense the available spectrum in real-time intelligently [85], [86]. ML plays a pivotal role in enabling CR capabilities in NFC systems, allowing devices to autonomously switch between frequency bands based on spectrum availability and quality. ML algorithms, such as SVMs and recurrent neural networks (RNNs), can implement CR functionalities in NFC devices. For example, an SVM-based classifier can predict the availability of specific frequency bands in the near-field region based on historical spectrum usage data [87]. Similarly, an RNN can model temporal dependencies in spectrum occupancy patterns to anticipate future availability, enabling NFC devices to select the best spectrum bands [88] dynamically. Moreover, ML-based CR facilitates spectrum sharing between multiple NFC systems in close proximity by dynamically adjusting transmission parameters (e.g., power levels, modulation schemes) to avoid interference. Additionally, federated learning (FL) can be used to train ML models for solving the decentralized spectrum sharing across multiple NFC devices [89].

### C. Interference Management

NFC generally operates over short distances and is widely employed for secure authentication and data exchange. However, interference from nearby NFC devices, electromagnetic noise, and multipath effects can significantly degrade communication performance. Traditional interference management relies on fixed model-based techniques, which lack adaptability in dynamic environments. ML provides intelligent, real-time solutions for interference detection, mitigation, and optimization in NFC networks, addressing the unique challenges of near-field communication. This subsection covers interference prediction, beamforming optimization, interference suppression, and ISAC.

### 1) Interference Prediction

ML models, such as RNNs and their variants (i.e., long short-term memory (LSTM) and gated recurrent units), can be used to predict interference patterns in NFC. These models can learn the temporal dependencies from historical data and environmental factors, enabling them to anticipate interference occurrences. For instance, the RNN application is explored to analyze co-channel interference on time-series data, capturing patterns such as periodic interference from nearby devices or noise spikes caused by environmental factors. By predicting interference, ML models enable proactive strategies, such as reconfiguring system parameters (e.g., transmission power, frequency bands) or adjusting communication schedules to minimize the impact of anticipated interference [90]. This capability is especially valuable in NFC, where interference can vary rapidly due to the short-range nature of communication and the presence of multiple devices in dense environments.

### 2) Beamforming Optimization

Beamforming plays a crucial role in NFC, as it aims to direct the signal toward the intended object and minimize interference. Traditional beamforming methods rely on predefined rules and static configurations, which are ineffective in near-field scenarios with rapidly changing channel conditions. The advanced DNN models are applied to optimize beamforming strategies in real-time by learning from accurate CSI and environmental data. Thus, DNNs are applied to determine the optimal beamforming weights based on real-time channel conditions, ensuring precise signal direction towards the intended receiver and reducing interference in near-field scenarios. Thus, the near-field beam focusing concept is designed to produce a spotlight-like pattern rather than the conventional flashlight-like pattern [91]. This intelligent beamforming approach is particularly beneficial in NFC, where precise signal directionality is essential for reliable communication over short distances.

### 3) Interference Suppression

Interference suppression (IS) is another critical area in the NFC system where ML demonstrates significant advantages over traditional methods. Traditional approaches often require domain expertise to manually extract features from interfering signals, which is less effective and time-consuming, especially in complex dynamic environments. On the other hand, DL models can automatically learn the characteristics of interfering signals from raw data, enabling more accurate and efficient IS. For example, DL models can distinguish between desired signals and interference, allowing them to filter out unwanted noise dynamically [92]. DL models can adapt to new interference sources by learning from real-time data, ensuring robust performance in dynamic environments. This capability improves communication quality and reduces the need for manual intervention, making NFC systems more autonomous and adaptable.

### 4) Integrated Sensing and Communication

ISAC, which functions in near-field scenarios, can enhance interference management. By utilizing the spherical wavefront characteristics of near-field propagation, ML approaches have the capability to achieve more accurate environmental sensing and inform interference mitigation strategies. For instance, the ISAC-enabled system is proposed for use in near-field scenarios to detect obstacles and adjust transmission parameters. ML models can be employed to sense data and allocate resources (e.g., frequency bands and time slots) more efficiently, reducing interference and improving overall system performance. This integration allows NFC systems to operate more efficiently in complex environments with constantly changing interference sources and channel conditions [93].

ML models can adapt to dynamic environments by learning from real-time data, enabling proactive interference management. This means that the ML models can solve complex tasks, such as interference prediction, beamforming optimization, and suppression, thereby minimizing the need for manual configuration and domain expertise. Moreover, ML techniques directly enhance key performance metrics, such as SINR and communication reliability, which are critical for NFC applications. By leveraging ML, NFC systems can achieve more robust and efficient communication, even in dense and dynamic environments.

### D. Energy Efficient Communication

Optimizing RA, particularly energy efficiency, is a critical concern in NFC systems in the context of 6G networks, where the deployment of ELAA, increasing complexity of communication scenarios, and integration of RIS demand intelligent solutions. ML plays a pivotal role in optimizing energy usage by enabling adaptive and data-driven decision-making. This subsection explains some of the key ways in which ML contributes to energy-efficient NFC systems.

### 1) Resource Allocation Optimization

RA is a fundamental challenge in NFC systems, where limited resources such as computational capacity, bandwidth, and transmit power must be utilized wisely and efficiently. ML approaches, particularly RL algorithms, are effective in this domain due to their ability to learn optimal policies through interaction with the environment. For instance, RL models can dynamically adjust bandwidth and transmit power levels based on real-time channel conditions, network congestion, and user demands. RL minimizes energy consumption while maintaining high-quality communication by learning from historical data and adapting to changing environments [94]. For example, RL-based power allocation strategies are demonstrated in RIS-assisted systems and achieve up to 44.9% energy savings compared to traditional methods [95].

### 2) RIS Configuration Optimization

RIS is a key enabling technology for future wireless communication, especially for 6G NFC, as RIS manipulates EM waves to enhance signal strength and coverage. However, it is always challenging to configure RIS elements (e.g., phase shifts and reflection angles) to achieve optimal performance while minimizing energy consumption. Recently, DRL has emerged as a powerful tool for optimizing RIS configurations. DRL models can learn to adjust the phase shifts of RIS elements in real time, ensuring that signals are directed toward users with minimal energy loss. For instance, a proposed DRL-based approach achieves significant EE improvements by reducing the power required for signal reflection while maintaining high SNR at the receiver [96].

### 3) Energy-efficient Beamforming and Signal Processing

Recently, ML techniques have been widely applied to optimize beamforming and signal processing in NFC systems. For example, DL models can design energy-efficient beamforming strategies that focus signals precisely on the intended receivers, reducing unnecessary energy dissipation [97]. ML-based signal processing algorithms can compress data and reduce computational overhead, enhancing energy efficiency. In [98], a proposed ML-based beamforming approach achieved a 30-50% reduction in energy consumption compared to conventional methods.

### 4) Predictive Maintenance and Energy Consumption Forecasting

Predictive Maintenance (PM) is a proactive maintenance strategy that uses data analysis to monitor equipment conditions, predict failures, and optimize performance. PM can enhance real-time monitoring, improve failure prediction accuracy, and maximize energy consumption when combined with ML models in NFC systems. ML models such as LSTM networks can predict future energy usage patterns and identify inefficiencies by analyzing historical energy consumption data. For example, LSTM networks can forecast the energy consumption of IoT devices connected via NFC and optimize their operation schedules to reduce energy waste [99]. This will not only extend the devices' battery life but also reduce the network's overall energy footprint. Moreover, in [84], LSTM-based PM has effectively reduced energy consumption in NFC-enabled IoT networks [100].

As highlighted above, incorporating ML into NFC systems holds significant potential to boost EE, promote sustainability, and reduce operational costs. These advancements are particularly crucial for 6G networks, where the proliferation of connected devices and the demand for high-speed, low-latency communication necessitate innovative energy-saving solutions.

## V. CASE STUDIES

In this section, we discuss the latest practical applications and experimental results of ML integration in NFC systems. More specifically, we focus on an RIS-assisted NFC system, which involves the BS, RIS, and end-user. Unlike FFC, where signals propagate over long distances with relatively stable channel conditions, near-field environments are characterized by rapidly changing channel dynamics, complex multipath effects, and strong spatial dependencies. These factors make traditional communication techniques less effective, necessitating the adoption of advanced methodologies to optimize performance. ML has emerged as a powerful tool to address these challenges, offering data-driven solutions that can adapt to the dynamic nature of near-field channels. In this section, we aim to highlight how ML approaches have been successfully applied in near-field communication to enhance signal reliability, optimize beamforming, mitigate interference, and improve channel estimation. These case studies not only showcase the current state-of-the-art but also provide insights into the practical challenges and future directions for integrating ML into NFC systems.

### A. Recent Advancements

In [101], ELAAs are studied in the new near-field beam management. Comprehensive numerical results are presented to validate the effectiveness of ELAA in near-field beam training designs. For this reason, perfect CSI beamforming is considered, assuming the BS is equipped with 512 antennas and the carrier frequency and transmit power are f = 100 GHz and 30 dBm, respectively. The reliability of the beam training success is evaluated based on finding the best beam codeword from the polar-domain codebook. The results are compared with those from an exhaustive search of near-field and two-phase near-field beam training. Based on the results, the proposed perfect CSI beamforming achieves a higher achievable rate than the other approaches. Perfect CSI beamforming exploits accurate channel state information to optimize beam selection, maximizing the achievable rate.

In [102], two approaches are analyzed for NFC in ultra-dense IoT applications. First, high Q coils are employed, including transmit, receive, and tag coils. Secondly, relay coils are utilized to extend the communication range. The mutual coupling effect is also evident in the close proximity of devices in ultra-dense IoT scenarios. The numerical results indicate that the distance between two tags should be approximately equal to the coil radius. Furthermore, it is demonstrated that utilizing high-Q coils and the relay coil can significantly enhance the reliability of NFC systems, achieving a range of 0.9–1.3 m for transmit powers ranging from 1 W to 10 W.

A two-stage beam training scheme for RIS-assisted near-field networks is proposed to achieve separate alignment in the angular and distance domains [38]. An AO algorithm is proposed to perform joint optimization iteratively, maximizing the achievable rate while reducing complexity. Specifically, the RIS coefficient matrix is optimized through the beam training process. The optimal combining matrix is obtained from the closed-form solution for the mean square error (MSE) minimization problem. The active beamforming matrix is optimized by exploiting the relationship between the achievable rate and MSE.

In [103], a CNN-based ML model is investigated for predicting near-field antenna performance parameters efficiently. The proposed method demonstrates the efficiency of the widely used rectangular path antenna. Furthermore, it has also been shown that the proposed CNN-based method is capable of predicting all necessary antenna parameters, including radiation pattern, resonant frequency, and directivity, with remarkable accuracy and improved system performance.

In [104], a DL-based beam training approach is proposed for a near-field MIMO system to reduce the training overhead and maximize the achievable rate. The angle and distance for the near-field are estimated by designing two neural networks. The proposed work is compared with the original schemes. The original scheme estimates the optimal near-field codeword directly based on the output of the neural networks, whilst the proposed scheme performs additional beam testing. Simulation results demonstrate that the proposed DL-based schemes can significantly enhance the achievable rate and reduce training overhead.

In [105], a DNN is trained for the near-field MIMO networks. The distance and angular information are obtained from a specific codebook, and the optimal distance and angle from the BS to the intended user are jointly predicted instead of being calculated independently. Moreover, two schemes, deep learning-based near-field beam training (DNBT) and DNBT with supplementary codewords (DNBT-SC), are proposed to improve the performance of beam training. The simulation results show that the proposed schemes can not only achieve better beam training performance but also maximize the achievable sum rate than the existing schemes.

In [106], a DL-based transmission architecture is proposed for RIS-aided near-field MIMO systems. This work studies a channel estimation scheme with low pilot overhead and a robust beamforming scheme. More specifically, an end-to-end (E2E) DL-based channel estimation framework is developed, comprising CSI feedback, pilot design, channel extrapolation, and subchannel estimation. The DL-based scheme is designed to simultaneously optimize the RIS phase and hybrid beamforming to maximize the sum rate of all UEs under imperfect CSI. In this study, numerical results are presented to demonstrate that the proposed DL scheme significantly outperforms traditional schemes in terms of reconstruction performance while imposing a notably reduced pilot overhead. Moreover, the proposed DL scheme also demonstrates superior beamforming designs in terms of achievable sum rate performance and robustness to imperfect CSI.

1n [107], a near-field beam training scheme is proposed based on the DL model. Moreover, a CNN is employed to strategically design the padding and kernel size that effectively extracts the complex CSI features. The loss function is assumed to be a negative value for the total achievable rate during the network training process, ensuring the amplitude constraint of the beamforming vector through a customized Tanh layer. In the end, simulation results demonstrate that the CNN-based model performs better than traditional approaches, does not rely on predefined codebooks, and only requires the estimated CSI as input to obtain the optimal beamforming vector.

## VI. CHALLENGES AND OPPORTUNITIES

The future of wireless communication depends on deploying ELAA and leveraging ML-driven intelligent decision-making for enhanced efficiency and connectivity. Integrating ML into NFC for future wireless communication presents both opportunities and challenges. NFC is expected to play a crucial role in 6G wireless communication networks, particularly for applications that require low latency, ultra-reliability, and energy-efficient communication. However, the dynamic nature of connected devices in near-field environments poses significant challenges, including energy optimization, real-time channel estimation, security vulnerabilities, and increased network complexity. Meanwhile, the advancement of ML presents unprecedented opportunities to tackle these challenges by offering real-time optimization, enabling intelligent beamforming, and adaptive antenna design of communication parameters. Leveraging advanced ML models for NFC can significantly enhance system performance, improve security, and boost scalability, paving the way for innovative applications in ultra-high-speed data transfer, wearables, and IoT. We explore some of the key research directions to highlight the transformative potential of ML in NFC for 6G as follows:

### A. Key Research Directions

The future of wireless communication depends on deploying ELAA and leveraging ML-driven intelligent decision-making for enhanced efficiency and connectivity. Integrating ML into NFC for future wireless communication presents a unique set of opportunities and challenges. NFC is expected to play a crucial role in 6G wireless communication networks, particularly for applications that require low latency, ultra-reliability, and energy-efficient communication. However, the dynamic nature of connected devices in near-field environments poses significant challenges, including energy optimization, real-time channel estimation, security vulnerabilities, and increased network complexity. Meanwhile, the advancement of ML presents unprecedented opportunities to tackle these challenges by offering real-time optimization, enabling intelligent beamforming, and adaptive antenna design of communication parameters. Leveraging advanced ML models for NFC can significantly enhance system performance, improve security, and boost scalability, paving the way for innovative applications in ultra-high-speed data transfer, wearables, and IoT. We explore some of the key research directions to highlight the transformative potential of ML in NFC for 6G as follows:

### 1) Enhanced Beamforming and Antenna Design

One of the appealing applications of ML in NFC for 6G is optimizing beamforming techniques and antenna design. Traditional beamforming methods, such as dynamic programming and AO, often struggle to adapt to the unique characteristics of NFC, such as spatial variations and rapid signal decay. ML algorithms, particularly DL-based approaches, can optimize beamforming patterns in real-time, ensuring precise control of EM fields in the near-field region. Moreover, intelligent antenna arrays can be designed to dynamically adjust their configurations based on environmental conditions and user proximity, thereby improving reliability and signal strength. This approach not only enhances the performance of NFC systems but also enables their deployment in complex and dynamic environments, such as smart cities and industrial IoT.

### 2) Energy Efficiency Optimization

EE is always a fundamental consideration for designing wireless communication and has become critical for NFC-enabled devices, especially in IoT and wearable applications with limited power resources. ML techniques, such as DRL and DL, can be employed to optimize energy consumption in NFC systems. These algorithms can dynamically adjust signal processing parameters to optimize power transfer, thereby minimizing energy usage while maintaining optimal performance. Furthermore, ML-driven approaches can enhance the efficiency of wireless power transfer in near-field regions, enabling the development of sustainable and long-lasting NFC devices. This emphasis on EE aligns with the broader goals of 6G networks for sustainability and green communication.

### 3) Real-Time Channel Estimation and Optimization

The highly dynamic nature of the connected device in NFC environments necessitates advanced techniques for real-time channel estimation and optimization. ML models, such as RNNs and CNNs, can predict and mitigate channel impairments, including interference and multipath fading. These models can also facilitate efficient feedback mechanisms for CSI, ensuring seamless communication in highly mobile and heterogeneous networks. By integrating ML for channel estimation processes, NFC systems can enhance reliability, reduce latency, and intelligently adapt to changing communication conditions. These enhancements are crucial for 6G applications, such as augmented reality (AR) and autonomous vehicles, where URLLC is indispensable.

### 4) Integration of Terahertz and Visible Light Communication

Integrating NFC with emerging technologies such as THz and visible light communication (VLC) represents a significant opportunity for 6G networks. ML-driven approaches can be employed to address challenges related to interference management and signal propagation in hybrid communication systems. For example, ML algorithms can optimize signal coordination and RA between NFC, THz, and VLC technologies, ensuring ultra-high-speed data transfer and seamless connectivity. This integration enables new possibilities for applications such as holographic communication, immersive virtual reality, and high-density IoT networks. By leveraging ML, NFC can play a pivotal role in converging diverse communication technologies and driving innovation in 6G networks.

*5) Privacy and Security*

Ensuring privacy and robust security becomes paramount as NFC becomes deeply integrated into 6G communication. ML-based intrusion detection systems can be developed to identify and mitigate threats like spoofing and eavesdropping attacks. Additionally, FL techniques can be leveraged to enable secure and privacy-preserving data sharing in NFC systems, minimizing the chances of exposure to external threats. Leveraging ML to NFC can ensure robust security mechanisms, safeguarding sensitive information while maintaining the efficiency and scalability required for 6G applications. This emphasis on security is particularly significant for applications such as healthcare, mobile payments, and innovative infrastructure, where user privacy and data integrity are of paramount importance. The explained key research directions are summarized in Fig. 9.

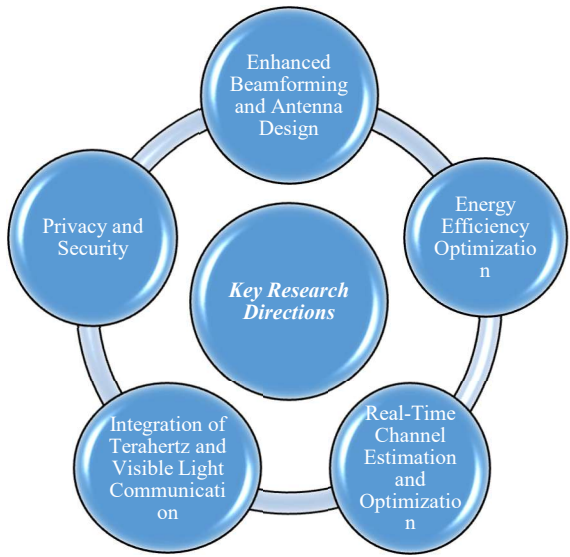

Fig. 9. Key research directions

*B. Challenges and Open Issues*

The integration of ML into NFC offers a transformative opportunity to enable advanced applications for 6G wireless communication systems, such as URLLC, energy-efficient IoT networks, and massive device connectivity. However, this integration presents significant challenges that must be addressed to realize its full potential. These challenges span standardization, operational, and technical domains, requiring interdisciplinary efforts from policymakers, stakeholders, and researchers. The following section discusses some of the key challenges and open issues.

*1) Near-field Propagation Models*

**Challenges:**

NFC operates in a regime where the EM fields behave differently from FFC. The near-field region is characterized by non-linear, complex, and spatially varying EM interactions that are highly sensitive to factors such as orientation, distance, material properties, and environmental conditions. These interactions are governed by Maxwell's equations, which necessitate a high computational cost to solve the problem in real time and are challenging to approximate accurately using simplified models. Some of the key challenges for near-field complexity propagation are summarized as:

• **Non-uniform field distribution:** The EM waves propagate in NFC involve rapidly decaying, highly localized fields. This makes it very challenging for the NFC to develop generalized models that accurately predict signal behavior across diverse scenarios.

• **Environmental Sensitivity:** The performance of NFC is highly influenced by environmental factors, including signal absorption, interference from nearby objects, and multipath effects. These factors introduce additional complexity to propagation modeling.

• **ML Models Dependency:** ML algorithms highly depend on accurate physical models for training and interpretation. However, oversimplified or inaccurate propagation models can lead to poor generalization, thereby limiting the performance of ML-driven NFC applications in real-world settings.

**Open Issues:**

Innovative solutions are necessary to address the challenges mentioned above. Some of the key open issues are:

• **Hybrid Modeling Approaches:** The traditional model relies on a physics-based approach (computationally expensive) and ML-driven approximation (lacks interpretability). One of the main concerns is how to effectively develop hybrid models that combine physics-based EM simulations with ML data-driven techniques to improve accuracy while reducing computational complexity. For example, incorporating the physical domain knowledge into NN architectures (e.g., physics-informed neural networks).

• **Real-Time Conditions Adaptation:** The propagation models must be designed to adapt to dynamic near-field environmental conditions, including varying interference, object movement, and dynamic material properties. The challenge is to develop a model that can efficiently learn and update its parameters in real-time. Techniques such as RL could be an adaptive propagation modeling approach for such a problem.

• **Benchmarking and Evaluation:** Despite advances in near-field propagation models, standard benchmarks and validation methods are still a question mark. Therefore, establishing well-defined datasets and performance metrics is critical to comparing different modeling approaches and ensuring the reliability of ML-based NFC systems.

### 2) Latency and Computational Overhead

**Challenges:**

NFC applications in 6G systems, such as real-time industrial automation (IA), AR, and autonomous systems, require high computational efficiency and ultra-low latency. However, deploying ML models in resource-constrained NFC devices presents significant challenges due to the resource constraints of low-power devices. Some of the challenges are discussed below:

• **Real-Time Inference**: ML models, particularly DL models, often require substantial computational resources, which may not be feasible for handling complex tasks, such as channel estimation or object detection. However, the processing capabilities of NFC devices are often limited, making real-time inference impractical.

• **EE Trade-offs**: NFC-enabled devices, particularly those in IoT or wearable applications, are often battery-operated and subject to strict energy constraints. Running computationally intensive ML models on such devices can lead to excessive energy consumption, thereby reducing the operational lifetime.

• **Hardware Limitations**: Many NFC devices lack specialized hardware (e.g., GPUs or TPUs) to accelerate ML computations, increasing latency and computational challenges.

**Open Issues:**

Several issues remain unresolved, hindering the reduction of latency and computational overhead despite advances in ML and hardware optimization. Some of the open issues are:

• **Optimized ML Models:** Most of the ML models are designed for high-performance computing environments rather than resource-constrained NFC devices. There is a need for ML models that are explicitly optimized for low-latency and low-power NFC applications.

• **Trade-off Between Complexity and Performance:** Techniques like quantization, model pruning, and knowledge distillation can reduce computational demands, but they often lead to performance degradation. Balancing efficiency and model accuracy remains a key issue in the NFC system using ML models.

• **Distributed Computing:** Offloading computations to cloud servers or nearby edge devices using edge computing or FL introduces security concerns and communication delays. Ensuring seamless, low-latency task offloading without compromising data privacy remains an unresolved issue.

### 3) Data Scarcity for ML Models

**Challenges:**

The success of ML algorithms in NFC applications depends on the availability of high-quality, diverse, and representative datasets. However, collecting such datasets is a significant challenge due to the unique characteristics of NFC scenarios.

• **Limited Real Scenario Data:** NFC applications often involve specialized use cases, such as smart manufacturing, IoT, or healthcare, where collecting large-scale real-time data is impractical, expensive, or time-consuming. For example, capturing data for NFC-enabled medical implants or industrial sensors may require controlled environments and specialized equipment.

• **Synthetic Data Limitations:** Although synthetic data generation can be used to address data scarcity, it often fails to capture the full complexity and variability of the real-world environment. Additionally, synthetic data may lack realistic environmental effects, interference, or noise, resulting in biased or overfitted ML models.

• **Privacy and Security Concerns:** NFC applications, such as those in healthcare or finance, may contain sensitive information, making it challenging to share or utilize for training ML models.

**Open Issues:**

• **High-Fidelity Data Generation**: Efficiently designing advanced simulation tools and generative models (e.g., GANs) to create high-fidelity synthetic datasets that closely mimic real-world NFC scenarios.

• **Federated Learning**: Leveraging FL techniques to train ML models on distributed datasets without compromising data privacy. This approach could enable collaborative model training across multiple devices or organizations.

• **Open Benchmarks and Datasets**: To standardize benchmarks and generate/share open datasets for NFC applications to facilitate research and development in this field.

### 4) Interoperability and Standardization

The lack of proper frameworks and standardized protocols for ML-driven NFC systems poses significant challenges to widespread interoperability and adoption.

**Challenges:**

• **Fragmented Ecosystem**: Without standardized protocols, different developers and vendors may implement proprietary ML solutions, leading to compatibility issues and hindering the integration of NFC into 6G networks.

• **Performance Evaluation:** The absence of standardized metrics and evaluation methodologies makes it difficult to compare the performance of different ML algorithms and NFC implementations.

• **Interoperability Challenges:** Ensuring that ML-driven NFC systems can operate seamlessly across diverse devices, platforms, and network architectures requires the development of universal standards that facilitate seamless integration.

**Open Issues:**

• **Industry-Wide Standards**: Establishing industry-wide standards for ML-driven NFC in 6G systems, covering data formats, communication protocols, and performance metrics.

• **Compliance and Certification**: Ensuring compliance, introducing certification programs with standardized protocols, and promoting interoperability across devices and networks.

• **Collaborative Frameworks:** Encouraging collaboration among industry stakeholders, standardization bodies, and

academic researchers to develop open frameworks for ML-driven NFC.

### 5) Interference and Coexistence with FFC

**Challenges:**

6G networks are expected to support both near-field and FFC simultaneously, creating potential interference and coexistence challenges.

• **EM Interference:** The proximity of near-field and FFC systems can lead to EM interference, degrading the performance of both systems. For example, NFC signals may interfere with far-field millimeter-wave or terahertz communication in 6G networks.

• **Dynamic Environments:** In dynamic environments, such as urban areas or industrial settings, the coexistence of near-field and FFC becomes even more challenging due to rapidly changing interference patterns and channel conditions.

• **Resource Allocation:** Efficiently allocating resources (e.g., spectrum, power) between near-field and FFC systems is a complex optimization problem that requires advanced ML techniques.

**Open Issues:**

• **Interference Mitigation Techniques:** Designing an advanced ML-based interference cancellation approach that can dynamically adapt to changing interference patterns. For example, using DRL to optimize RA in real-time problems.

**Cross-Layer Optimization:** Exploring cross-layer optimization approaches that jointly optimize higher-layer protocols (i.e., routing, scheduling) and physical layer parameters (i.e., power control, beamforming) to enhance coexistence.

**Coexistence Protocols:** Developing robust protocols ensuring seamless operation of NFC and FFC in distinct environments. Advanced ML models can be effective in predicting and mitigating interference scenarios.

These challenges and open issues are summarized in Fig. 10 for better and easier understanding.

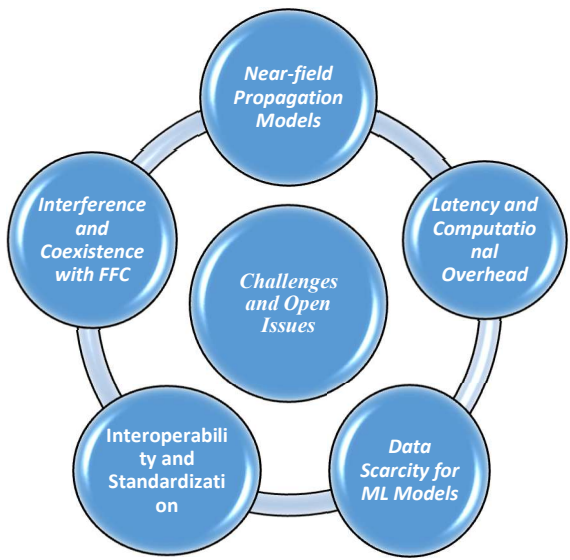

Fig. 10. Summary of challenges and open issues

### C. Opportunities for Future Research

The integration of ML in NFC for the upcoming wireless communication networks (i.e., 6G) opens up a wide array of research opportunities that align with the vision of sustainable, intelligent, and adaptive communication systems. Some of the possible key future research directions are:

### 1) ML for URLCC

Latency and reliability play a crucial role in designing a wireless communication network. In NFC, achieving these (i.e., latency and reliability, commonly known as URLLC) is crucial for mission-critical applications such as autonomous vehicles, industrial automation, and remote surgery. In the future, advanced ML models, such as DRL and GNN, should be explored to meet demanding requirements (i.e., latency of sub-1ms and reliability of 99.9999%). Moreover, edge computing and ML integration could further enhance URLLC performance by enabling localized decision-making and reducing latency.

### 2) AI-driven Self-organizing Networks

ML development based on self-organizing networks (SONs) for NFC can revolutionize network management by enabling autonomous optimization of parameters such as RA beamforming and power levels. One potential future research direction is to focus on scalable and real-time ML approaches, such as RL and FL, to address the challenges associated with heterogeneous networks and dynamic environments. Furthermore, SONs could enable self-sustaining networks in smart cities and industrial IoT applications by integrating with 6G's AI-native architecture.

### 3) Cross-Layer Optimization

The integration of ML across multiple layers of communication stacks (media access control (MAC), physical, and network layers) can improve NFC performance and efficiency. In future research, frameworks such as multi-agent RL (MARL) and TL should be leveraged to address cross-layer dependencies and trade-offs. Simulation tools and testbeds (hardware implementation) are essential for validating these approaches and ensuring they are compatible with the unified network architecture of 6G.

### 4) Green NFC

The objective of the 6G network is to achieve sustainable communication by developing energy-efficient ML algorithms for NFC. This is a promising research direction. In the future, novel techniques such as spiking neural networks (SNN) and quantum ML should be explored to optimize energy consumption and enable energy harvesting in NFC systems. Furthermore, green NFC must be aligned with international standards and regulations to reduce the environmental impact of communication networks.

### 5) Human-Centric NFC Applications

With the aid of ML-enabled NFC, customized and context-aware services can be delivered, such as smart healthcare, AR, and immersive entertainment. In future research, user behavior data should be investigated to enhance NFC capabilities while addressing security and privacy concerns simultaneously. Moreover, to design truly human-centric NFC systems that prioritize accessibility and user experience, collaboration is crucial with fields such as human-computer interaction (HCI) and psychology.

## VII. CONCLUSION

In this paper, we have provided a comprehensive overview of the role of ML in NFC for 6G networks. We first briefly discussed the background of NFC, providing the theoretical and mathematical explanation to differentiate between NFC and FFC. We shed light on a real-world problem (RIS-assisted Near Field MIMO communication) to enhance our understanding. We then provide a detailed review of the role of ML in NFC, describing different approaches and applications of ML in NFC. We outlined some of the relevant case studies to highlight the effectiveness of ML in NFC. We also identified key research directions, challenges, and opportunities for ML-enabled NFC for 6G networks. Our survey reveals that advanced ML approaches are crucial for achieving wavefront behavior in NFC, enabling optimal operation in dynamic and unpredictable environments.

Laboratory, Centre for advanced Simulation and Visualization (V-Sim). He is Editor in Chief of SIMULATION (Sage) and a member of the editorial board of the IEEE Computing in Science & Engineering, Wireless Networks (Elsevier), and The Journal of Defense Modeling and Simulation (SCS). He is a fellow of the Society for Modeling and Simulation International (SCS).

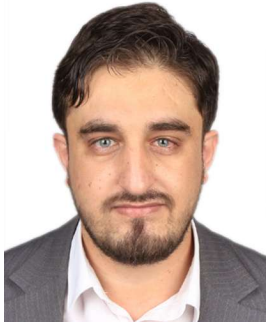

**Amjad Iqbal** is currently a Postdoctoral Research Fellow in the Department of Systems and Computer Engineering at Carleton University, Canada. He received his B.Sc. degree in Electrical Engineering from CECOS University, Peshawar, Pakistan, in 2014, and his M.Sc. degree in Electrical and Electronic Engineering from the University of Engineering and Technology (UET), Peshawar, in 2017. In 2022, he earned his Ph.D. in Wireless Communication from Universiti Tunku Abdul Rahman (UTAR), Malaysia.

He is a highly skilled researcher with a strong passion for advancing the field of wireless communication. His research interests include resource allocation optimization, machine learning, deep reinforcement learning, 5G/B5G cellular networks, and reconfigurable intelligent surfaces (RIS). He has published several research papers in top-tier journals and prestigious international conferences in these domains. In addition to his research contributions, he also serves as a reviewer for leading journals and conferences in the field. As a Postdoctoral Research Fellow, he continues to make impactful contributions to next-generation wireless communication systems.

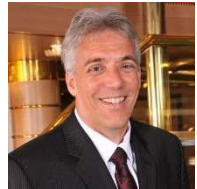

**GARY BOUDREAU** (Senior Member, IEEE) received the B.A.Sc. degree in electrical engineering from the University of Ottawa, in 1983, the M.A.Sc. degree in electrical engineering from Queens University, in 1984, and the Ph.D. degree in electrical engineering from Carleton University, in 1989. From 1984 to 1989, he was a Communications Systems Engineer with Canadian Astronautics Ltd., and from 1990 to 1993, he was a Satellite Systems Engineer with MPR Teltech Ltd. From 1993 to 2009, he was with Nortel Networks in a variety of wireless systems and management roles within the CDMA and LTE basestation product groups. In 2010, he joined Ericsson Canada, where he is currently the Director of RAN Architecture and Performance with the North American CTO Office. His research interests include digital and wireless communications, signal processing, and machine learning.

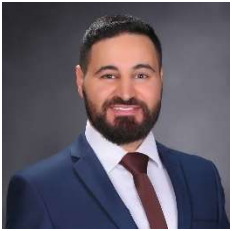

**ALA'A AL-HABASHNA** received his Master of Engineering degree from Memorial University of Newfoundland in 2010 and his PhD in Electrical and Computer Engineering from Carleton University in 2018. Currently, Dr. Al-Habashna is an Adjunct Research Professor at Carleton University, Ottawa, Canada, and an Assistant Professor at the School of Computing and Informatics, Al Hussein Technical University, Amman, Jordan. He has won multiple awards, including excellence and best paper awards. He has worked as a reviewer for many conferences and journals and as a Technical Program Committee Member of multiple conferences. His current research interests include 5G wireless networks, machine learning, computer vision, IoT applications, localization, multimedia communication over wireless networks, signal detection and classification, cognitive radio systems, and discrete-event modeling and simulation.

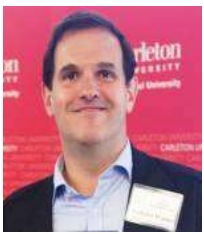

**GABRIEL WAINER** received the PhD degree (Highest Hons.) from UBA/Université d'Aix-Marseille III, Marseille, France, in 1998. He held visiting positions with the University of Arizona; LSIS (CNRS), Université Paul Cézanne, University of Nice, INRIA Sophia-Antipolis, Université de Bordeaux (France); UCM, UPC (Spain), University of Buenos Aires, National University of Rosario (Argentina) and others. He is currently a full professor with Carleton University, Ottawa, ON, Canada, where he is also the head with the Advanced Real-Time Simulation